\documentclass[a4paper,UKenglish,cleveref,autoref,thm-restate]{lipics-v2021}

\NewDocumentEnvironment{longversion}{ +b }{\iflongversion#1\fi}{}

\usepackage{todonotes}
\usepackage{booktabs}

\title{Trusting AI to increase productivity? Perspectives Across the Global North and South}

\titlerunning{Trusting AI to increase Productivity?} 

\author{Adam Bokun}{University of Bayreuth, Germany}{adam.bokun@uni-bayreuth.de}{0009-0006-0942-7969}{}
\author{Shalini Chakraborty}{University of Bayreuth, Germany}{shalini.chakraborty@uni-bayreuth.de}{0000-0002-9466-3766}{}

\authorrunning{Bokun and Chakraborty}
\Copyright{Adam Bokun and Shalini Chakraborty}



\nolinenumbers
\begin{document}

\maketitle

\begin{abstract}
Generative AI (GenAI) tools are widely used in academia and software development, where productivity gains may depend not only on technical capabilities but also on users' trust and contextual factors. This paper presents emerging results from an exploratory study investigating the relationship between trust in GenAI and perceived productivity, motivated by Global South contexts. We conducted a systematic literature review, complemented by a grey literature analysis and a survey study. The literature review identified no peer-reviewed evidence at the intersection of GenAI trust, productivity, and Global South settings, while the grey literature revealed only limited insights. At the time of writing, the survey has received 36 valid responses from participants across both the Global North and Global South, including individuals with cross-regional experiences. Preliminary results suggest that respondents born and working in the Global South tended to trust AI more, but did not usually report clear productivity gains from using it. In contrast, respondents born and working outside the Global South reported stronger productivity gains and greater time savings, even though they showed less trust in generative AI. These findings suggest that trusting AI is not enough on its own; productivity also depends on access, the type of task, and how much users need to check the output.


\end{abstract}

\section{Introduction}
\label{sec:introduction}
Generative artificial intelligence (GenAI) tools are increasingly embedded in academic and software development workflows, supporting activities such as writing, coding, data analysis, and research assistance \cite{brynjolfsson2025generative}. As a result, GenAI is often presented as a productivity-enhancing technology \cite{noy2023experimental,tabachnyk2026achieving}.
However, AI-driven productivity gains are not determined solely by the technical capabilities of large language models (LLMs). Rather, they are strongly influenced by user trust in these systems \cite{vossing2022designing}. Whether individuals adopt, rely on, and meaningfully integrate AI-assisted tools into their work depends on the extent to which they perceive these systems as reliable, useful, and beneficial~\cite{2025WhatGuidesOurChoises}. Understanding the relationship between trust and productivity is therefore essential for assessing the impact of GenAI across different contexts.
Existing research has examined trust in AI~\cite{2025RethinkTrustAI,2022TrustHumanAI}, AI adoption~\cite{2025AISoftwareEngineering}, and productivity-related outcomes~\cite{2025ToUseOrNotToUse,2025GeminiAtWork,2025ExaminingUseImpactAI} in human-AI interaction. However, much of the available evidence has been produced in contexts where access to digital infrastructure, paid AI tools, institutional support, and training is relatively stable. These assumptions may not hold equally in global contexts. In particular, there is limited systematic evidence on how academics and developers working in the Global South perceive and trust AI as a productivity-enhancing tool. Structural constraints such as unequal access to computational resources, data, training, and institutional support can shape both trust in AI and its realized productivity benefits. 

This creates an important empirical gap. While GenAI is increasingly discussed as a productivity-enhancing technology, less is known about how trust in GenAI relates to perceived productivity among users working under different infrastructural, economic, and institutional conditions. This gap matters for empirical software engineering because developers' use of AI tools is shaped not only by model performance, but also by the conditions under which these tools are accessed, evaluated, and trusted.



Building on this gap analysis, this paper presents emerging results from an exploratory study on trust and perceived productivity in AI-assisted work, with a particular focus on Global South contexts. 
We first conducted a systematic literature review (SLR) to investigate existing peer-reviewed research at the intersection of AI-assisted tools, trust, productivity, and the Global South. The review revealed little to no peer-reviewed evidence addressing this combined 
space, highlighting a substantial research gap. 
To broaden our understanding beyond academic publications, we subsequently conducted a grey literature review to identify ongoing discussions and contextual factors surrounding trust in AI, and productivity in diverse contexts.
At the third stage, based on the findings from the grey literature review, we designed a survey study aimed at understanding perceptions of trust and productivity in relation to GenAI use, particularly from a Global South perspective. However, to establish a broader contextual baseline and account for diverse experiences, we extended the study to participants from both the Global North and Global South, including individuals with cross-regional experiences. Rather than shifting the focus away from the Global South, this comparative perspective serves as a first step toward identifying patterns and contextual differences that can guide future investigations with a stronger and more targeted emphasis on Global South settings.
As data collection and analysis are still ongoing, this paper reports emerging results intended to stimulate discussion and provide early insights into a rapidly evolving research area.



\section{Background}
\label{sec:background}

GenAI tools are increasingly used in knowledge-intensive workflows, both academic and software development context. In software engineering, developers use GenAI tools for code generation, debugging, documentation, test generation, refactoring, and learning unfamiliar APIs or frameworks \cite{2026KnowledgePrompting}. In academic work, GenAI tools are used for writing support, literature exploration, summarization, explanation, data analysis, and research assistance \cite{2024UsingAIinAcademicWriting}. Although these tasks differ in their domain-specific requirements, they share common features: information processing, problem solving, and the production or evaluation of textual or technical outputs.
For this reason, academics and software developers provide a useful empirical context for studying GenAI-assisted productivity. In both groups, GenAI tools may reduce search effort, accelerate routine tasks, or support idea generation. At the same time, incorrect or misleading outputs can create additional verification work. Productivity therefore depends not only on whether GenAI can generate outputs quickly, but also on whether users can evaluate and integrate these outputs into their work.

Trust in AI systems can be understood as a user's willingness to rely on the system under conditions of uncertainty and potential risk \cite{2004TrustInAutomation}. In the context of GenAI, trust is closely related to a user's attitude that the system will help them achieve their goals under conditions of uncertainty and vulnerability \cite{2025WhatGuidesOurChoises}.
This distinction is important because GenAI tools can produce fluent but incorrect or incomplete outputs. As a result, trust does not necessarily mean accepting AI-generated content without verification. Instead, users may develop calibrated trust: they rely on GenAI for some tasks, verify outputs for others, and avoid using it in situations where errors would be costly.
The relationship between trust and productivity is not straightforward. On the one hand, trust may enable productivity gains because users are more likely to adopt GenAI tools, rely on their outputs, and integrate them into routine workflows. A developer who trusts an AI-generated code suggestion may complete a task faster \cite{2026DevsAndGenAI}, while an academic who trusts an AI-generated explanation may reduce time spent searching for introductory material \cite{2024UsingAIinAcademicWriting}.
However, insufficient or excessive trust may reduce productivity. If users do not trust the GenAI output, they may avoid using the tools or spend substantial time verifying even low-risk output. In contrast, if users trust AI outputs too much, they may accept incorrect suggestions, leading to errors, rework, or lower-quality results. Productivity gains from GenAI therefore depend on calibrated trust: enough trust to make the tool useful, but enough caution to verify outputs when necessary.

In this paper, productivity is treated as \emph{perceived productivity} rather than directly measured task performance. We focus on users' perceptions of time savings, reduced effort, improved work quality, and overall satisfaction with AI-assisted work \cite{noy2023experimental,brynjolfsson2025generative}.
The term \emph{Global South} is used as an analytical category describing regions that often face unequal access to economic, technological, and institutional resources within the global knowledge economy \cite{dados2012global}. We do not treat the Global South as a homogeneous context; countries, institutions, and user groups differ substantially in infrastructure, education, digital skills, language environments, and access to AI technologies \cite{ragnedda2020understanding}.
Nevertheless, the concept is useful for examining how structural conditions may shape GenAI adoption and use. Access to GenAI tools may depend on internet connectivity, electricity, paid subscriptions, institutional support, local language availability, and training opportunities \cite{holmes2023guidance}. These factors can influence both trust in AI systems and their perceived productivity benefits.

\section{Methodology}
\label{sec:methodology}

\subsection{Research Design and Search Strategy}
\label{subsec:design}

Our aim is to examine how trust in AI systems relates to perceived productivity in academic and software development work, with particular attention to Global South contexts. Because the topic combines several research areas that are often studied separately, we first conducted a systematic literature review. The purpose of this search was to identify peer-reviewed studies that explicitly connect trust in AI or GenAI systems with productivity-related outcomes, and that also consider Global South, developing-country, low-resource, or comparable contextual settings.
The search strategy was structured around three conceptually connected blocks. The first block captures AI and GenAI systems. The second block captures trust-related concepts, including trust and trustworthiness. The third block captures productivity-related outcomes, including productivity, performance, and efficiency. A fourth contextual block was added to identify studies that explicitly engage with Global South or comparable settings. The final search string was:

\begin{verbatim}
("artificial intelligence" OR AI OR "generative AI" OR GenAI)
AND (trust OR trustworthiness) AND (productivity OR performance OR efficiency)
AND ("software engineering") AND ("Global South" OR "developing countries" 
OR "developing regions" OR LMIC OR "low-income" OR Africa OR "South Africa" 
OR "Latin America" OR "Southeast Asia" OR India OR rural OR underserved 
OR marginalized OR "low-resource" OR "low-literate" OR "novice technology users")
\end{verbatim}

Based on the search string, Google Scholar returned 4{,}977 articles published between 2022 and 2025. This period was selected because the widespread adoption of contemporary GenAI tools in academic and software development contexts accelerated following the public release of large-scale conversational AI systems.
To identify studies relevant to our research focus, we applied a staged screening procedure (Table~\ref{tab:screening-process}). The objective was not to capture all work related to AI, trust, or productivity, but rather to identify studies that explicitly examined the relationship between these concepts in software engineering or related knowledge-work settings, with particular attention to Global South contexts.

The initial set of 4{,}977 papers was reduced to 3{,}180 after excluding papers with fewer than two citations or fewer than four pages. Restricting the results to conference proceedings yielded 182 papers, while limiting the dataset to CORE A and CORE B venues further reduced the set to 38 papers. Following full-text screening, none of the remaining papers satisfied the inclusion criteria. Although a substantial body of literature exists on AI, trust, and productivity individually, we found no academic studies directly investigating the relationship between trust in GenAI and productivity from the perspective considered in this review.

\begin{table}[t]
\centering
\caption{Overview of the academic literature screening process.}
\label{tab:screening-process}
\begin{tabular}{lr}
\hline
Screening stage & Number of papers \\
\hline
Initial Google Scholar results & 4{,}977 \\
After excluding papers with fewer than two citations or fewer than four pages & 3{,}180 \\
After retaining papers published in conference proceedings & 182 \\
After retaining CORE~A and CORE~B conferences & 38 \\
After full-text screening & 0 \\
\hline
\end{tabular}
\end{table}

\subsection{Grey Literature Research}
As the SLR did not identify any studies that satisfied all inclusion criteria, we complemented it with a grey literature review. This additional review aimed to capture emerging evidence from reports, policy documents, and industry analyses, recognizing that research on GenAI is evolving rapidly and that relevant insights may appear in non-academic sources before reaching peer-reviewed venues.
We identified grey literature through three targeted Google searches. The first search used a structured query with Boolean-style keyword combinations: \textit{("artificial intelligence" OR "generative AI" OR GenAI) (trust OR trustworthiness) (productivity OR performance OR efficiency) "software engineering" ("Global South" OR "developing countries" OR Africa OR "Latin America" OR India OR "Southeast Asia") filetype:pdf}. The second and third searches used broader queries: \textit{GenAI trust productivity in Global South} and \textit{GenAI trust productivity in Global South software engineering}. After removing duplicate and overlapping results, an initial set of \textbf{19 sources} was obtained. These sources were then screened for credibility and relevance. To be included, a source had to originate from a recognized organization and address at least three of the following themes: GenAI, productivity, trust, software or IT work, and contextual barriers affecting adoption in the Global South. The final dataset consisted of reports and analyses published by organizations such as MIT Technology Review Insights, McKinsey, OECD, Globalization Partners, and EY India. The selected sources represent a range of evidence types, including executive surveys, expert interviews, and economic and policy analyses.

The reviewed sources suggest that GenAI is widely framed as a productivity-enhancing technology, but that its benefits may be unevenly distributed. McKinsey estimates potential productivity gains from GenAI across sectors, including corporate IT and product development~\cite{mckinsey2023economic}. EY India similarly highlights productivity opportunities in India, especially in software development, BPO services, and IT consulting~\cite{ey2025india}. However, these sources mainly estimate potential gains rather than explaining how users' trust shapes those gains.
The grey literature also points to unequal conditions for realizing AI-related productivity gains. The OECD argues that AI may contribute to productivity growth, but that adoption and preparedness vary substantially across countries, with low-income and lower-middle-income countries facing weaker readiness~\cite{oecd2025productivitydivide}. This is directly relevant to the Global South focus of this study, because it suggests that productivity gains from GenAI cannot be assumed to occur uniformly across contexts.
Trust appears less consistently than productivity in the reviewed grey literature. MIT Technology Review Insights and Globalization Partners discuss trust-related concerns in GenAI adoption, but mainly treat trust as a practical adoption issue rather than as a clearly defined construct~\cite{gp2025aiatwork,mit2023greatacceleration}. Overall, the grey literature supports the need for an explicit empirical focus on how trust, productivity, and access conditions interact in GenAI-assisted work.
\subsection{Survey Design}
Following the SLR and grey literature searches, we identified the need to move beyond existing evidence and capture current perspectives and experiences regarding trust and productivity in GenAI use. While the study was initially motivated by understanding these dynamics within Global South contexts, narrowing participation exclusively to Global South audiences at this stage was not considered efficient for obtaining a sufficiently broad and diverse sample. Moreover, increasing mobility across regions means that many individuals originating from the Global South currently work in the Global North and vice versa. To capture this diversity and establish an initial baseline understanding, we therefore adopted a comparative framework to first examine broader patterns before moving toward more focused Global South investigations.

We designed an exploratory survey that examines how users perceive the relationship between trust in GenAI and productivity, and whether this relationship is shaped by contextual factors such as access, cost, infrastructure, training, and institutional support.

The target population includes academics, researchers, and software developers. The survey was designed and administered using the university's survey platform following prior ethics approval. It was distributed globally through professional networks, mailing lists, and social media channels. Respondents were asked to indicate both their region of origin and current working region, enabling exploratory comparative analyses and the identification of contextual factors that may inform future research with a stronger focus on Global South settings.
The survey consists of four main parts:
\begin{romanenumerate}
\item The first part collects demographic and contextual information, including respondents' role, primary working region, and type of work.
\item The second part asks about GenAI usage, including frequency of use and the tasks for which respondents use GenAI tools.
\item The third part measures trust in GenAI. Trust-related items were adapted from exploratory study on trust in GenAI among students, which surveyed students' trust in GenAI tools and their perceptions of how GenAI affects performance in computer science courses \cite{2024TrustAmongStudents}.
\item The fourth part measures perceived productivity. Productivity-related items were adapted from study on measuring productivity and trust in human--AI collaboration \cite{2024TakeItLeaveIt}.
\end{romanenumerate}

The survey also examines contextual barriers and enabling factors identified in the grey literature review, including access costs, internet reliability, institutional support, training opportunities, language limitations, and the ability to verify AI-generated outputs. These factors may influence how effectively users adopt and benefit from GenAI tools.
%

\subsection{Data Analysis}
Survey responses were analyzed using descriptive and exploratory methods. Most questions use Likert-scale responses to assess perceptions of trust, productivity, and access conditions. Likert-scale items were coded from $-2$ to $2$, where $-2$ indicates strong disagreement, $0$ indicates a neutral response, and $2$ indicates strong agreement. Higher values therefore indicate stronger agreement with the corresponding trust, productivity, or access-related statement. Responses with insufficient information for the main constructs, including trust, perceived productivity, and working region, were excluded from the main analysis.

For each respondent, we calculated an aggregate trust score and an aggregate perceived productivity score by averaging the corresponding survey items. These aggregate scores were used to summarize overall patterns in trust and productivity perceptions. Because the current survey sample is small ($n=36$), especially for respondents born in the Global South and currently working in the Global South ($n=3$), subgroup comparisons are reported descriptively and are not interpreted as statistically robust evidence of group differences.
Respondents were grouped according to their Global South background and current working context: born in the Global South and currently working in the Global South, born in the Global South and currently working outside the Global South; finally, born outside the Global South and currently working outside the Global South.
We report counts, averages, and descriptive patterns for these groups. Reported access barriers and time-saving responses were summarized using counts and percentages.

Open-ended responses were analyzed thematically to contextualize the quantitative results. The first author conducted the initial quantitative analysis, calculated the aggregate scores, and assigned initial codes to the open-ended responses. The second author rechecked the quantitative results, reviewed the coding decisions, and examined whether the interpretation was consistent with the underlying data. Disagreements were resolved through discussion.

\section{Results}
\label{sec:results}

\subsection{Survey respondents}
\label{subsec:survey-respondents}

The survey received 36 valid responses. Of these, 3 respondents were born in the Global South and currently work in the Global South, 11 were born in the Global South and currently work outside the Global South, and 22 were born outside the Global South and currently work outside the Global South. Because the first subgroup is very small, the subgroup comparisons are interpreted descriptively and should not be read as statistically robust evidence of group differences (Table \ref{tab:respondent-groups}).

\begin{table}[htbp]
\centering
\caption{Respondent groups by Global South background and current working context.}
\label{tab:respondent-groups}
\begin{tabular}{lr}
\hline
Respondent group & Number of respondents \\
\hline
Born in the Global South; currently working in the Global South & 3 \\
Born in the Global South; currently working outside the Global South & 11 \\
Born outside the Global South; currently working outside the Global South & 22 \\
\hline
Total & 36 \\
\hline
\end{tabular}
\end{table}

\subsection{Trust and productivity}
\label{subsec:trust-productivity-results}

\begin{figure}[htbp]
\centering
\includegraphics[width=0.6\textwidth]{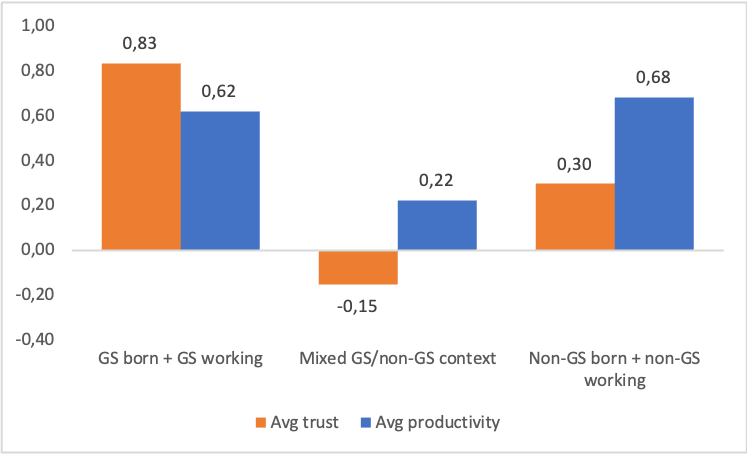}
\caption{Average trust and perceived productivity scores by geographic context. Trust scores are based on six survey items and productivity scores on seven survey items. Likert-scale answers were coded from $-2$ to $2$, where higher values indicate stronger agreement.}
\label{fig:avg-trust-productivity}
\end{figure}

Likert-scale items were coded from $-2$ to $2$, where $-2$ indicates strong disagreement, $0$ indicates a neutral response, and $2$ indicates strong agreement. For each respondent, we first calculated an individual trust score by averaging the six trust-related items and an individual productivity score by averaging the seven productivity-related items. We then calculated the mean trust and productivity scores for each geographic group. Across all respondents, the average trust score was $0.20$ and the average productivity score was $0.54$. Figure~\ref{fig:avg-trust-productivity} presents the average trust and perceived productivity scores by geographic context. Respondents born and currently working in the Global South had an average trust score of $0.83$ and an average productivity score of $0.62$. Respondents in the mixed Global South/non--Global South context had an average trust score of $-0.15$ and an average productivity score of $0.22$. Respondents born and currently working outside the Global South had an average trust score of $0.30$ and an average productivity score of $0.68$.

Respondents were asked how much time they usually save per task when using GenAI tools. The response options were converted into approximate minute values: no time saved was coded as 0 minutes, less than 5 minutes as 5 minutes, 5--15 minutes as 10 minutes, 15--30 minutes as 22.5 minutes, and more than 30 minutes as 30 minutes. The ``more than 30 minutes'' category was coded as 30 minutes, making this estimate conservative. We then calculated the average reported time saved for each geographic group. As shown in Table~\ref{tab:saved-time}, respondents born and currently working in the Global South reported an average of 10.83 minutes saved per task, respondents in the mixed Global South/non--Global South context reported 12.27 minutes, and respondents born and currently working outside the Global South reported 21.48 minutes.

\begin{table}[htbp]
\centering
\caption{Average estimated time saved per task when using GenAI tools.}
\label{tab:saved-time}
\begin{tabular}{lr}
\hline
Geographic context & Average time saved per task (minutes) \\
\hline
GS born + GS working & 10.83 \\
Mixed GS/non-GS context & 12.27 \\
Non-GS born + non-GS working & 21.48 \\
\hline
\end{tabular}
\end{table}

\begin{figure}[htbp]
\centering
\includegraphics[width=0.9\textwidth]{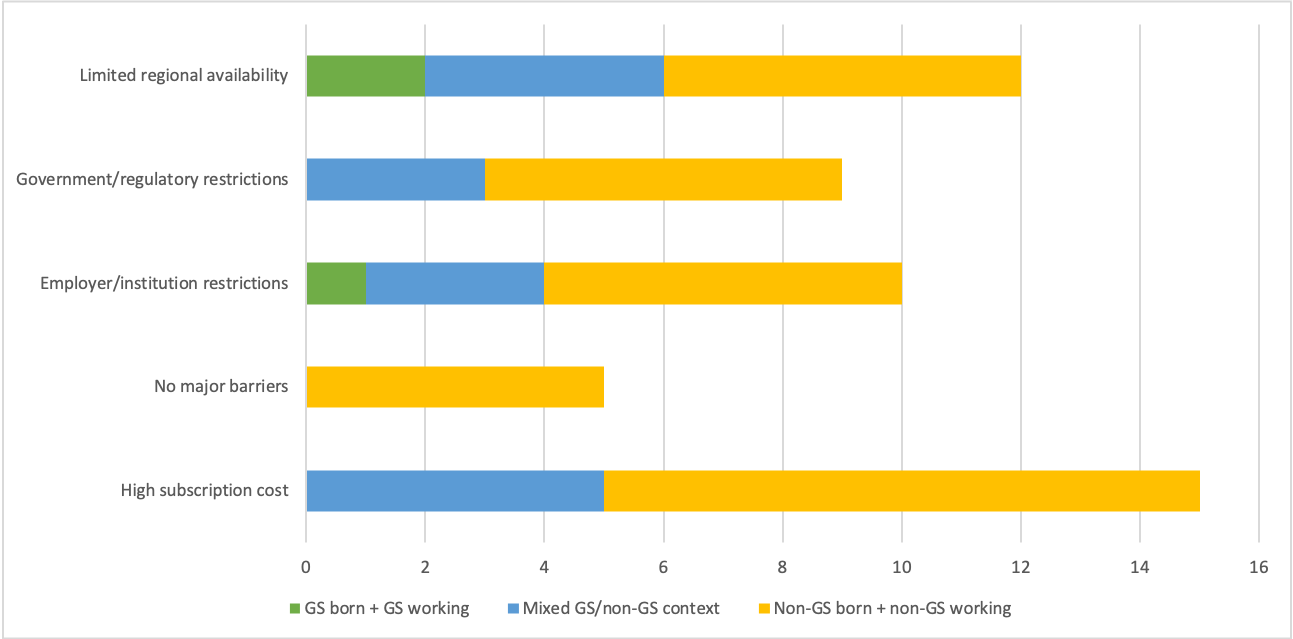}
\caption{Reported barriers to accessing AI tools for work by geographic context. Respondents could select multiple barriers.}
\label{fig:barriers-context}
\end{figure}

Overall, 33\% of respondents born and currently working in the Global South, 45\% of respondents in the mixed Global South/non--Global South context, and 50\% of respondents born and currently working outside the Global South reported at least one access restriction. Figure~\ref{fig:barriers-context} presents the reported barriers to accessing AI tools for work by geographic context. The most frequently reported barrier was high subscription cost, followed by limited regional availability, employer or institutional restrictions, and government or regulatory restrictions. Some respondents also reported no major barriers to accessing AI tools for work.

Respondents were asked whether they think trusting AI increases their productivity. Overall, 27 out of 36 respondents answered yes, corresponding to 75\% of the sample. Table~\ref{tab:trust-increases-productivity} reports the response counts by geographic context. Because subgroup sizes are small and uneven, especially for respondents born and working in the Global South, the geographic breakdown is reported as counts rather than interpreted as a stable cross-regional pattern.

\begin{table}[htbp]
\centering
\caption{Responses to whether trusting AI increases productivity.}
\label{tab:trust-increases-productivity}
\begin{tabular}{lrrr}
\hline
Geographic context & No & Yes & Total \\
\hline
GS born + GS working & 0 & 3 & 3 \\
Mixed GS/non-GS context & 6 & 5 & 11 \\
Non-GS born + non-GS working & 3 & 19 & 22 \\
\hline
Total & 9 & 27 & 36 \\
\hline
\end{tabular}
\end{table}

\subsection{Open-Ended Responses}
\label{subsec:open-ended-results}
The open-ended responses were coded into descriptive categories to summarize how respondents explained their trust, mistrust, concerns, and perceived productivity gains related to GenAI. Because some responses contained more than one idea, the counts in Table~\ref{tab:open-ended-categories} refer to theme mentions rather than individual respondents. The most frequent category was productivity and time saving, including faster work, routine task support, and delegation of simple tasks. Illustrative responses included descriptions of GenAI as ``a second pair of eyes,'' as a tool that helps users ``deliver faster,'' and as a system whose outputs may still require checking because models are ``prone to hallucination.''

\begin{table}[ht]
\centering
\caption{Descriptive categories from open-ended responses.}
\label{tab:open-ended-categories}
\begin{tabular}{p{0.38\textwidth}rp{0.42\textwidth}}
\toprule
Category & Mentions & Example quote \\
\midrule
Productivity and time saving & 14 
& ``It speeds up the work. Teams can deliver faster and iterate over the product roadmap.'' \\

Human verification and supervision & 9 
& ``I use it mostly as a second pair of eyes, I don't trust it unless I was already happy with the results.'' \\

Reliability and output quality concerns & 9 
& ``Output sometimes lacks quality. Risks of missing/publishing a bug.'' \\

Limited context and task complexity & 7 
& ``They hallucinate and it's impossible to provide them all the context.'' \\

Learning, research, and starting-point support & 5 
& ``It often allows me to have some sort of starting point quickly from where to start my work from.'' \\

Data, confidentiality, and bias concerns & 4 
& ``The results can be biased, skewed or compliant towards local geopolitical norms.'' \\
\bottomrule
\end{tabular}
\end{table}

\section{Discussion}
\label{sec:discussion}

%
Despite thousands of papers touching on AI, trust, productivity, or Global South contexts, our structured search did not identify a single peer-reviewed study that explicitly examined all three dimensions together. This does not suggest that the topics are unimportant; rather, it indicates a fragmentation of the research landscape. Trust is frequently studied as a determinant of AI adoption, productivity is often discussed independently of trust, and Global South research tends to focus on access, infrastructure, or adoption challenges. The intersection of these themes remains largely unexplored.
The grey literature review helped contextualize this gap. Industry and policy sources commonly describe GenAI as a productivity-enhancing technology, but often focus on expected economic or organizational benefits rather than on how trust shapes productivity. Sources also point to uneven AI readiness and access conditions across countries. This is important for the present study because productivity gains from GenAI may depend not only on whether users trust the system, but also on whether they have stable access to tools and the ability to verify AI-generated output.

The preliminary survey results show a contrast between trust and perceived productivity across geographic contexts. Respondents born and currently working in the Global South reported the highest average trust score ($0.83$), while respondents born and currently working outside the Global South reported a lower average trust score ($0.30$). However, the productivity results show a different pattern: respondents born and working outside the Global South reported the highest average productivity score ($0.68$) and the greatest estimated time savings, while respondents born and working in the Global South reported a slightly lower productivity score ($0.62$). The mixed Global South/non--Global South group reported the lowest average trust score ($-0.15$) and the lowest average productivity score ($0.22$). Taken together, these results suggest that higher trust does not automatically translate into stronger productivity gains. Given the small and uneven subgroup sizes, especially for respondents born and working in the Global South, these patterns should be read as descriptive indications rather than statistically robust group differences.

One possible explanation is that users working outside the Global South may benefit from conditions that make GenAI easier to integrate into everyday work: better access to paid tools, more institutional support, stronger workplace adoption, or more opportunities to use AI in professional workflows. In this sense, respondents who move from Global South to non--Global South contexts may be adapting to working conditions shaped by the Global North, but their experiences may remain distinct from respondents born and working outside the Global South. The mixed-context group is therefore important because it may capture transitional experiences between different access, training, and workplace environments.
Accessibility remains central to this interpretation. Although access restrictions were reported in all groups, the reported barriers show that subscription cost, regional availability, and institutional or regulatory restrictions may shape how GenAI tools are used. These barriers can affect productivity even when users trust AI. For example, a user may trust a tool but still be unable to use the most capable version, may lack institutional permission to use it for work, or may need additional time to verify outputs because of limited training or support. 

The open-ended responses provide additional insight into how participants perceive the relationship between trust and productivity. Many respondents associated productivity gains with support for routine tasks, faster completion of work, debugging assistance, learning, and the generation of initial drafts or starting points. At the same time, concerns about output quality, hallucinations, limited contextual understanding, confidentiality, and bias appeared repeatedly throughout the responses. 
One participant summarized this relationship particularly well:
\emph{``I don't have to trust the system for it to increase my productivity. I cross check for errors, at least currently as I don't use a paid service and use basic models.''}
This statement highlights an interesting interaction between trust, productivity, and accessibility. Even without complete trust in the system, the participant still perceives productivity benefits. At the same time, the reference to using only freely available models suggests that access to more advanced tools may influence both trust and perceived effectiveness.
A second participant offered a different perspective:
\emph{``In my experience, it is more like delegating and reducing mental effort and menial tasks but then we always have to verify the results and information, so overall productivity is the same but effort required is reduced.''}
This response raises an important question regarding how productivity itself is understood by users. While some participants appear to define productivity in terms of time savings or task completion, others associate it with reduced cognitive effort, lower mental workload, or the delegation of routine activities. These responses suggest that perceived productivity is a multidimensional concept and that future analyses should distinguish between productivity as speed, productivity as output, and productivity as reduced effort.

\section{Conclusion}
\label{sec:conclusion}

This paper presents emerging results from an ongoing study of trust in GenAI and perceived productivity. While our literature review identified no peer-reviewed studies examining trust, productivity, and Global South perspectives together, the grey literature and preliminary survey findings suggest that these dimensions are closely connected. Notably, higher trust did not necessarily correspond to higher perceived productivity, indicating that contextual factors such as access, verification effort, and working environment may play an important role. Although the current sample is small and unevenly distributed, the findings highlight the need for further empirical investigation of how trust, productivity, and accessibility interact across different contexts. Future work will expand the dataset, particularly from Global South populations, and refine our understanding of these relationships.

\newpage
\section*{Data Availability}

The anonymized survey data, survey instrument, coding scheme, and supplementary analysis materials are available on \href{https://zenodo.org/records/20431971}{\textbf{Zenodo}}. To protect participant privacy, free-text responses were anonymized before publication, and any identifying information was removed.

\bibliographystyle{plainurl}
\bibliography{refs}

\end{document}